\documentclass[superscriptaddress,showkeys,twocolumn,nofootinbib,longbibliography]{revtex4-1}

\usepackage{amsmath}
\usepackage{amssymb}
\usepackage{graphicx}
\usepackage{epsf}
\usepackage{slashed}
\usepackage{enumitem}
\usepackage[czech,english]{babel}
\usepackage[cp1250]{inputenc}
\usepackage{hyperref}
\usepackage{xcolor}

\usepackage[only,llbracket,rrbracket,llparenthesis,rrparenthesis]{stmaryrd} 
\usepackage{accsupp} 

\newcommand{\I}{\ensuremath{\mathrm{i}}}
\newcommand{\e}{\ensuremath{\mathrm{e}}}
\renewcommand{\d}{\ensuremath{\mathrm{d}}}
\newcommand{\cc}{\ensuremath{\mathrm{c.c.}}}

\newcommand{\angstrom}{\mbox{\normalfont\AA}}
\newcommand{\eV}{\ensuremath{\mathrm{\,eV}}}

\newcommand{\amu}{\ensuremath{\mathrm{\,a.m.u.}}} 

\newcommand{\abs}[1]{\left|#1\right|}

\usepackage{bbm} 

\usepackage[nodayofweek]{datetime}
\newdateformat{mydate}{\twodigit{\THEDAY}{ }\shortmonthname[\THEMONTH], \THEYEAR}
\usepackage[normalem]{ulem}

\usepackage{color}
\usepackage{ulem} 

\definecolor{myred}{rgb}{1,0,0}
\definecolor{mygreen}{rgb}{0,0.8,0.2}
\definecolor{myblue}{rgb}{0,0,1}

\definecolor{Ared}{rgb}{1,0.7,0}
\definecolor{Agreen}{rgb}{0.7,0.8,0.2}
\definecolor{Ablue}{rgb}{0,0.7,1}

\renewcommand{\emph}[1]{\textit{#1}}

\begin{document}

\title{Solitons in Peyrard--Bishop model of DNA and the Renormalization group method}

\author{Filip Blaschke}
\email{filip.blaschke@fpf.slu.cz}
\affiliation{Research Centre of Theoretical Physics and Astrophysics, Institute of Physics, Silesian University in Opava,\\ Bezru\v{c}ovo n\'am\v{e}st\'i~1150/13, 746~01 Opava, Czech Republic}
\affiliation{Institute of Experimental and Applied Physics, Czech Technical University in Prague,\\ Husova~240/5, 110~00 Prague~1, Czech Republic}

\author{Ond\v{r}ej Nicolas Karp\'i\v{s}ek}
\email{karponius@gmail.com}
\affiliation{Institute of Physics, Silesian University in Opava,\\ Bezru\v{c}ovo n\'am\v{e}st\'i~1150/13, 746~01 Opava, Czech Republic}

\author{Petr Bene\v{s}}
\email{petr.benes@utef.cvut.cz}
\affiliation{Institute of Experimental and Applied Physics, Czech Technical University in Prague,\\ Husova~240/5, 110~00 Prague~1, Czech Republic}

\begin{abstract}
We investigate solitons in Peyrard--Bishop model of DNA molecule using Renormalization group methods which provide a systematic way for perturbation analysis. Small amplitude expansion is carried out in both the continuous and discrete limits. We review exact solution for the continuous model. Further, we discuss reliability of the solitonic solutions and argue that the envelope should propagate with a group velocity contrary to previous proposals.
\end{abstract}

\keywords{Solitons; Peyrard--Bishop model; Exact solutions; BPS limit; DNA, Renormalization group}

\maketitle


\section{Introduction \& Results}


The Deoxyribo-Nucleic Acid (DNA) molecule is the Nature's instrument of proliferating information about biological organisms across geological time scales. In this regard, DNA must accommodate two opposing functions: protection of genetic information from the environment and facilitating rapid, reliable and repeatable access to that information. From this point of view, it seems that the dynamics of DNA must be particularly fine-tuned.

As is well known, genetic information is encoded in a sequence of pairs of nitrogenous bases held together via hydrogen bonds: adenine-thymine (A-T) and guanine-cytosine (G-C). Each base is protected by a sugar-phosphate group forming together a nucleotide, while these nucleotides are arranged in two antiparallel strands of polynucleotides in a right-handed double-helix structure.

Due to the complexity of DNA molecule -- roughly $100$ degrees of freedom per base pair with, typically, $10^{10}$ bases per strand \cite{Gaeta} -- it is practically impossible to analyse its dynamics from the first principles, e.g., using all atoms quantum mechanics. It is therefore of practical necessity to develop and study simplified descriptions of DNA, where but the most important effective degrees of freedom are kept. In particular, mechanical models of DNA with only a single effective degree of freedom per base are nowadays very popular. Among these, most prominent ones are a two-rod model by Ludmila V.~Yakushevich \cite{Yak}, where only the torsion modes of the double helix are studied. Another one is the model of Peyrard and Bishop \cite{Bishop} (PB) where the focus is on transverse modes between the nucleotides. In both approaches, the effective theories which emerges are one-dimensional and they support solitonic solutions.

The importance of solitons for understanding processes of DNA transcription, replication or gene expression has been recognised long time ago \cite{Englander}. For instance, the impact of solitons for the formation of denaturation bubbles (precursors for protein synthesis) has been studied extensively \cite{Tabi1, Dauxois} (also see references in the review paper \cite{Slobodan}).

In \cite{Yak} Yakushevich introduced a hierarchy of models describing torsion modes of DNA which in the continuous limit reduce to either sine-Gordon (sG) model or some derivations of it. As it is well known, sG model supports solitonic solutions  whose existence is guaranteed by topology. Further, this theory is known to be integrable (in the sense that its Hamiltonian can be expressed in terms of action-angle variables) and even the solutions describing scattering of arbitrary many solitons are known in analytic form.

In contrast, in the PB model there is no topology to make the existence of solitons manifest. Rather, they come about through interplay between dispersion and self-focusing non-linear effects. Indeed, the behaviour of small amplitude waves in the PB model is dispersive in the sense that different wavelengths travel with different velocity leading to the gradual disintegration of wave packets. On the other hands, non-linear terms, such as cubic and quartic terms in the wave's amplitude, leads to self-focusing. 
When these effects are balanced a stable (or long-lived) solitonic waves becomes the effective modes of energy transfer.

The traditional way of analytic demonstration of existence of solitons has been to utilize multiple-scale perturbation expansion method \cite{Slobodan}, which is a very useful tool for constructing global approximate solutions to non-linear differential equations. In this method, a naive perturbation expansion is carried out until the so-called secular terms are encountered. These terms quickly outgrow all other terms making the perturbation series divergent. To remedy this, amplitude of the zero order solution (typically a monochromatic wave) is promoted to a function of a set of scaled variables, which are introduced in such a way as to remove all secular terms in the perturbation expansion to a given order. In this way, the amplitude becomes a function of the coordinates, transforming the zero order solution into a global approximate solution.  

There are two main drawbacks of this method. First, the correct choice of scaled variables is not obvious and can be only justified \emph{a posteriori}. Second, going to the next order is not straight-forward as there might be a need for new scales, which are impossible to predict in advance. Thus, the mastery of the multiple-scale method requires a lot of previous experience, intuition and -- for the lack of a better word -- art. 

In case of the PB model, the multiple-scale analysis requires the promoted amplitude to satisfy Non-linear Schr\"odinger (NLS) equation as a condition for cancellation of secular terms. It is exactly NLS equation which has solitonic solutions and which is -- like sG equation -- fully integrable. 

This procedure has become a standard theoretical argument for showing the presence of solitons in the PB model and models related to it. However, there are difficulties on both technical and conceptual level. To engage in multiple-scale analysis to begin with one must assume a continuous limit for the scaled variables in sharp contrast to discrete nature of the underlying model. This further restricts the region of validity of the resulting approximate solution. Secondly, the choice of scaled variables is guessed without any physical justification, which raises doubt about uniqueness. Perhaps a different choice of scaled variables would lead to new or more general type of solitonic solutions? Lastly, as emphasized above and which is particularly true for PB-like models, multiple-scale seems to be just too cumbersome a technique for systematic analysis of solitons on DNA precisely because it is hard to go beyond leading order.
 
In this paper, we show that employing a different, more general method of analysis, one can eliminate all the above objections. In turn, we present a systematic and straight-forward technique of analysing solitons in mechanical models of DNA which is free of any doubt about the uniqueness of its solutions or their region of validity. 
 
The key ingredient is the Renormalization group (RG) method for finding global approximate solutions of differential equations as developed by Chen, Goldenfeld and Oono \cite{Oono1}. It was devised precisely to overcome many drawbacks of more traditional perturbation methods (such as those described in now a classic textbook \cite{Bender}). In fact, in \cite{Oono2}, these authors demonstrates that RG method provides a unifying frame for many particular techniques, such as multiple-scale, boundary layers, asymptotic matching, WKB and others.



In this paper we use RG methods to rederive (and clarify) some of the old results about solitons in PB model in both continuous and discrete limits. In former case, we also recall all of the exact static solutions. In Sec.~\ref{sec:II} we first illustrate RG method on a simple example of a non-linear oscillator. Sec.~\ref{sec:III} serves as introduction into the PB model, while in Sec.~\ref{sec:IV} we analyse its continuous limit. The application of RG method in the continuous limit is rather revealing. We find that solitonic solution of the NLS equation is correct small-amplitude approximation to the full equations of motion only in a small interval of velocities centered around the group velocity of the carrier wave. Furthermore, solitons moving exactly at group velocity recover Lorentz covariance -- a true symmetry for continuous PB model -- despite it being lost at the level of NLS equation. 

These findings are also reflected in the discrete case, which we tackle in Sec.~\ref{sec:V}. At the leading order of the RG perturbation expansion, we find the same effective NLS equation for the envelope as reported in other studies \cite{Slobodan}. But, again, we find that solitons should move with a group velocity of its carrier wave or very close to it. This contradicts the idea of a \emph{coherent mode} in which the envelope travels at \emph{phase} velocity of the carrier wave \cite{Zdravkovic2}. We discuss the relevance of this finding in Sec.~\ref{sec:VI}.

\section{RG-improved perturbation theory}
\label{sec:II}

Let us illustrate the workings of the RG method on a simple example.
Let us consider a particular anharmonic oscillator governed by the equation
\begin{equation}
\label{eq:quartosc}
\ddot y+y+\varepsilon y^3 = 0 \,,
\end{equation}
where $y$ is a distance out of equilibrium position and $\varepsilon$ a small positive parameter, i.e., $0< \varepsilon \ll 1$. Solving the naive perturbation series
\begin{equation}
y = y_0 + \varepsilon y_1 +\varepsilon^2 y_2 + \ldots
\end{equation}
up to the first order yields
\begin{eqnarray}
\label{eq:sol1}
y_{\mathrm{B}} &=&  A_0 \e^{\I t}+ \frac{\varepsilon}{8}A_0^3 \e^{3\I t} +\frac{3\I \varepsilon(t-t_0)}{2}A_0 \abs{A_0}^2\e^{\I t} + \cc\,,
\hspace{6mm}
\end{eqnarray}
where the subscript $_{\mathrm{B}}$ stands for `bare'. Here, $A_0$ is a complex parameter and $\cc$~denotes complex conjugation.

Notice that in the last term we have a secular term $(t-t_0) \e^{\I t}$ (`secular' means that it separates from other terms as $t \to \infty$). Its presence is a result of a resonance in the first order, indicating that the naive perturbation series breaks down. Indeed, this term would grow much larger than previous terms in the series for sufficiently large $|t-t_0|$. 
Roughly speaking, the perturbation series breaks down when the first order becomes as important as zero order, that is $\varepsilon |t-t_0| \sim 1$. In other words, we can rely on the perturbative solution only within the range
\begin{equation}
t_0 - \frac{1}{\varepsilon} 
\ \lesssim \  
t 
\ \lesssim \ 
t_0 + \frac{1}{\varepsilon} \,.
\end{equation}

Of course, for arbitrary small $\varepsilon$ this range can be arbitrarily large. However, we can never take Eq.~\eqref{eq:sol1} as a good \emph{global} approximation. In fact, one can see very easily that since there is a conserved quantity (energy)
\begin{equation}
E = \frac{1}{2}\dot y^2 + \frac{1}{2}y^2 + \frac{\varepsilon}{3}y^3 + \frac{\varepsilon^2}{4}y^4 \,,
\end{equation}
the exact solution must be bounded for all $t$. A secular term clearly violates this property. It is often said that secular terms are artifacts of the perturbation series and if one were to sum all of them up, one would obtain a finite expression. While true, for most non-linear equations such approach would be of no use since it is nearly impossible to find explicit formula for secular terms to all orders. 

The way that RG method deals with secular terms is as follows. First, let us introduce the \emph{renormalization scale} $\tau$ by a benign shift $t_0 \to t_0 -\tau + \tau$. Now we redefine the 'bare' amplitude $A_0$ in terms of the 'dressed' amplitude $A$ in such a way that the dependence on $t_0$ disappears: 
\begin{equation}
A_0 = A\Bigl(1-\frac{3\I \varepsilon (t_0-\tau)}{2} \abs{A}^2 + \mathcal{O}(\varepsilon^2)\Bigr) \,.
\end{equation}
Thus, the solution \eqref{eq:sol1} transforms into
\begin{equation}
y = A \e^{\I t}+ \frac{\varepsilon}{3}A^3 \e^{3\I t} +\frac{3\I \varepsilon(t-\tau)}{2}A \abs{A}^2\e^{\I t} + \cc
\end{equation}

\begin{figure}[t]
\begin{center}
\includegraphics[width=1.0\columnwidth]{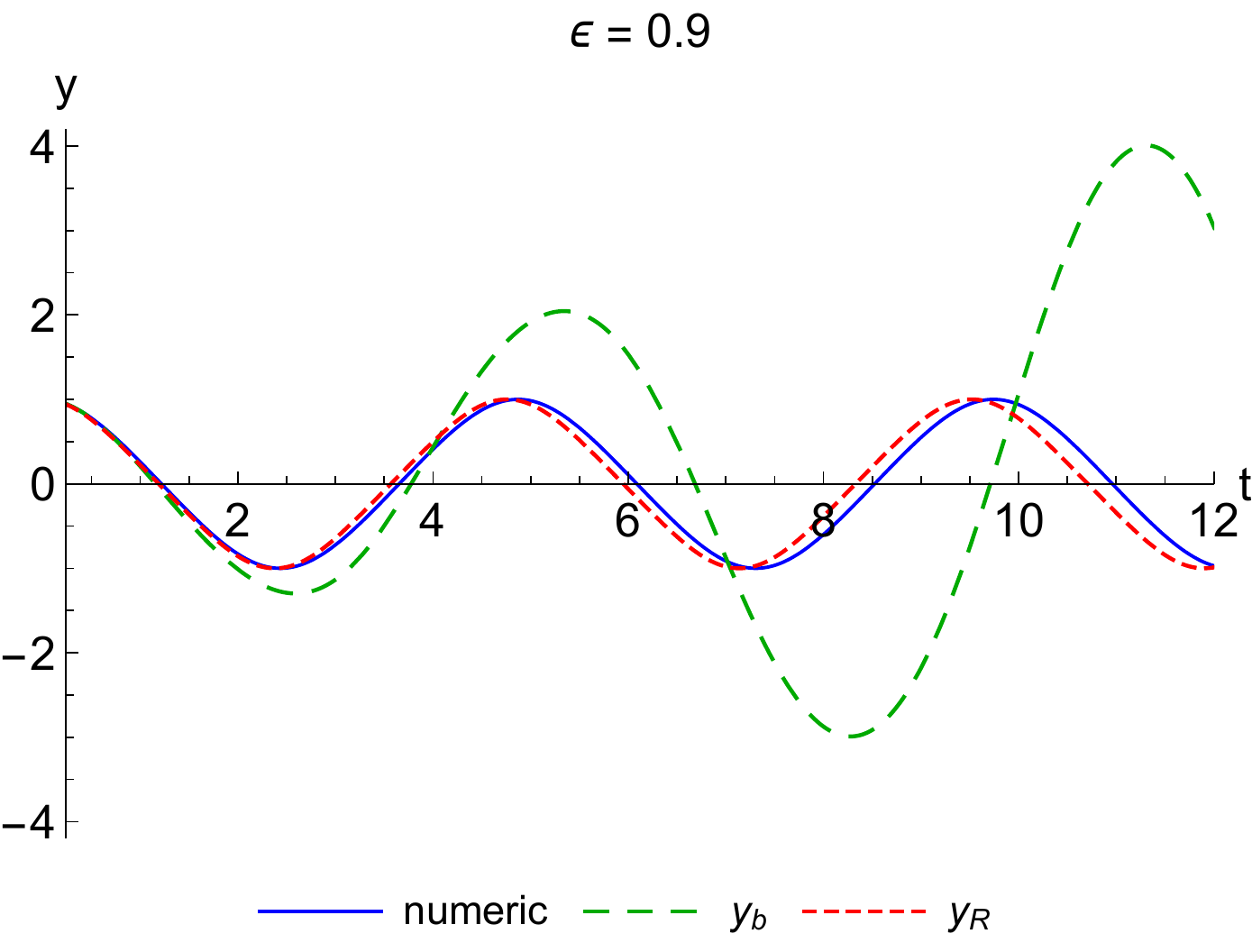}
\caption{\small Comparison among the exact (numerical) solution of Eq.~\eqref{eq:quartosc}, bare solution $y_{\mathrm{B}}$, Eq.~\eqref{eq:sol1}, and renormalized solution $y_{\mathrm{R}}$, Eq.~\eqref{eq:sol2}. We have chosen relatively large $\varepsilon = 0.9$ in order to make the exact and renormalized solutions distinguishable by eye.}
\label{fig:01}
\end{center}
\end{figure}

It seems that all that changed is notation, but conceptually we have leaped forward. Since $\tau$ is an artificial parameter, we can set it to whatever we want. A shrewd option is to set $\tau = t$, since it eliminates the secular term and save the perturbation series. But, we have to be careful to check that the solution $y$ \emph{does not depend on} $\tau$. In other words, we must demand that
\begin{equation}
\label{RGE}
\frac{\d y}{\d \tau} = 0
\hspace{5mm}
\mbox{for all $t$} \,.
\end{equation}
This equation is referred to as the RG equation. Since $A \equiv A(\tau)$ can be an arbitrary function of $\tau$, the RG equation \eqref{RGE} is recasts as
\begin{equation}
\partial_\tau A = \frac{3\I \varepsilon}{2}A \abs{A}^2+ \mathcal{O}(\varepsilon^2) \,,
\end{equation}
with the solution (to the first order in $\varepsilon$)
\begin{equation}
A = R \e^{\I \frac{3 R^2}{2}\varepsilon t} \,,
\end{equation}
where $R$ is an arbitrary real constant. To be specific, let us choose the initial conditions $y(0) = 1$, $\dot y(0) = 0$. The `renormalized' solution (hence the subscript $_{\mathrm{R}}$) then reads
\begin{eqnarray}
y_{\mathrm{R}} &=&  2 R \cos\left[\left(\frac{3 R^2}{2}\varepsilon +1\right)t\right ]
+ \frac{\varepsilon}{4}R^3 \cos\left[\left(\frac{3 R^2}{2}\varepsilon +1\right)3t\right]
\nonumber \\ && {}
+ \mathcal{O}(\varepsilon^2) \,,
\label{eq:sol2}
\end{eqnarray}
where $R = 1/2 - \varepsilon/64 + \mathcal{O}(\varepsilon^2)$ is a real root of $2R + R^3 \varepsilon / 4 - 1 = 0$. We compare this solution $y_{\mathrm{R}}$ and the `bare' solution $y_{\mathrm{B}}$, Eq.~\eqref{eq:sol1}, with exact (numerical) solution in Fig.~\ref{fig:01}.

\section{Model}
\label{sec:III}

Peyrard--Bishop (PB) model is an effective, mechanical model of DNA. The nucleotides (bases) are represented as points of the common mass $m$ forming two strands. Along the strands, they are coupled to the nearest neighbors via harmonic potential (representing the covalent bonds) while the strands themselves are linked via Morse potential (hydrogen bonds). Since the covalent bonds are far stronger than hydrogen bonds, only vibrational modes are considered in the PB model, i.e., the longitudinal and torison modes are ignored. In this way, the problem becomes one-dimensional.

Originally, PB model did not take into account the helicoidal structure of DNA \cite{Bishop}, but this is remedied in the so-called helicoidal Peyrard--Bishop model proposed later \cite{Dauxois}, which introduces additional harmonic potential between $n$-th base on one strand and $(n\pm h)$-th base on the other, where $h$ is usually taken to be 5 \cite{Slobodan}. In this way, the bases are coupled to the nearest neighbors across the helicoidal staircase. Other models have been considered. For example, one can also include the viscosity via friction force \cite{Tabi} or one can consider the so-called Peyrard--Bishop--Dauxois model of DNA \cite{Dauxois2}. Other references can be found in \cite{review}. In this paper, we will concern ourselves only with the PB model, but the results presented here can be easily extended to other models as well.

If we denote the displacement of the $n$-th base of the first strand from the equilibrium configuration as $u_n$ and the same for the second strand as $v_n$, we can write the Hamiltonian of the Peyrard--Bishop model as\footnote{We follow the notation of \cite{Slobodan}.}
\begin{eqnarray}
H &=& \sum_n \bigg\{ \frac{m}{2} \big(\dot u_n^2 + \dot v_n^2 \big) 
\nonumber \\ && \hspace{4.5mm} {} 
+ \frac{k}{2} \Big[ \big(u_n - u_{n-1} \big)^2 + \big(v_n - v_{n-1} \big)^2\Big]
\nonumber \\ && \hspace{4.5mm} {}
+D\Big[\e^{-a(u_n-v_n)}-1\Big]^2 \bigg\}
\,.
\end{eqnarray}
Here, $k$ is the string constant for the nearest neighbors along each strand and $m$ is the average mass of the base.
The parameters $D$ and $a$ are the depth and the inverse width of the Morse potential, respectively. In this paper, we adopt the following values (taken from \cite{Slobodan})
\begin{equation}
\begin{array}{c}
k = 0.74892 \displaystyle\frac{\eV}{\angstrom}\,, \hspace{3mm}
m = 307.2\, \amu\,, \\[8pt]
a = 1.2\, \angstrom^{-1}\,, \hspace{3mm}
D = 0.07 \eV\,, \hspace{3mm}
l = 3.4\, \angstrom\,,
\end{array}
\end{equation} 
where $l$ is the distance between neighboring bases which will be useful later. Furthermore, in these units one tick of the clock amounts to
\begin{equation}
\mathrm{t.u.} \equiv \angstrom \sqrt{\frac{\amu}{\eV}} \approx 1.024\times 10^{-14}\, \mathrm{s} = 102.4\, \mathrm{ps} \,.
\end{equation} 

An advantageous change of coordinates is
\begin{equation}
x_n \equiv \frac{u_n+v_n}{\sqrt{2}} \,, \hspace{5mm}
y_n \equiv \frac{u_n-v_n}{\sqrt{2}} \,,
\end{equation}
as it completely decouples in-phase and anti-phase motion, that is, $x_n$ representing common motion of both strands at the $n$-th base and $y_n$ representing mutual separation of strands at the $n$-th base. Of these two, only $y_n$ is influenced by the non-linear Morse potential and is therefore important for solitons.

The equations of motion are, respectively,
\begin{eqnarray}
m \ddot x_n &=& k\big(x_{n+1}+x_{n-1}-2x_n\big) \,, 
\\
\label{eq:y}
m \ddot y_n &=& k\big(y_{n+1}+y_{n-1}-2y_n\big)
\nonumber \\ && {}
+ 2\sqrt{2} a D \Big(\e^{-a\sqrt{2}y_n}-1\Big) \e^{-a\sqrt{2}y_n} \,.
\end{eqnarray}
The first equation of motion is linear and therefore completely solvable. The general solution is a linear combination of monochromatic waves of the form
\begin{equation}
x_n = A\,\e^{\I (nq l-\omega_a t)} + \cc \,,
\end{equation}
where $A$ denotes the complex amplitude, $q$ is the wave number, the parameter $l$ is, as we have already mentioned, the distance between two sites and $\omega_a$ is the acoustical (or phonon) frequency given as
\begin{equation}
\omega_a^2 = \frac{4k}{m} \sin^2\!\bigg(\frac{ql}{2}\bigg) \,.
\end{equation} 

In what follows, we will be concerned with solving the second equation \eqref{eq:y}. In the next section, we take a continuous limit to illustrate the power of RG method in the most simple setting.

\section{Lessons from the continuous limit}
\label{sec:IV}

Let us first investigate solitonic solutions of Eq.~\eqref{eq:y} in the continuous limit 
where the distance between sites $l$ is taken to zero while at the same time $l^2 k \equiv \tilde k$ is kept constant. Let us denote the continuous variable tracing the distance along strands as $nl \to x$ and the field variable which replaces transversal motion  at the $n$-th side as $y_n(t) \sim y(nl,t) \to y(x,t)$. 
The equation of motion \eqref{eq:y} becomes
\begin{equation}
m\partial_t^2 y - \tilde k \partial_x^2 y = 2\sqrt{2} a D \Big(\e^{-a\sqrt{2}y}-1\Big) \e^{-a\sqrt{2}y} \,.
\end{equation}
To simplify things, we switch to dimensionless coordinates $\tilde t \equiv \omega_g t $ and $\tilde x = \omega_g x\sqrt{m}/\sqrt{\tilde k}$ and rescale the field as $y = \tilde y/(a\sqrt{2})$. Dropping the $\tilde{}\,$ sign from all symbols for brevity, we arrive at the equation
\begin{equation}
\label{eq:y2}
\partial^2 y = \Big(\e^{-y}-1\Big) \e^{-y} \,,
\end{equation}
where we employed relativistic notation $\partial^2 \equiv \partial_\mu \partial^\mu = \partial_t^2 -\partial_x^2$. As we see, the continuous limit brought about enhancement of symmetry. We have started with a Newtonian mechanics of a system of particles and end up with a relativistic field theory in $(1+1)$-dimensions.
As we will show, this has non-trivial consequences on the solitonic solution of this equation. 

To simplify things even further we make a substitution $y = \log(1+u)$. We obtain
\begin{equation}
\label{eq:u}
\big(\partial^2 +1\big)u = (\partial_\mu u)(\partial^\mu u)- u \, \partial^2 u \,.
\end{equation}
This equation is non-linear only in second order in $u$. Moreover, it has a peculiar structure, namely, its right-hand side can be written as $-\frac{1}{2} \mathcal{D}^2 \big(u\cdot u\big)$,
where $\mathcal{D}$ denotes the Hirota derivative, i.e.,
\begin{eqnarray}
\mathcal{D}^2 \big(u\cdot u\big) &\equiv&
\partial_\xi^2 \Big(u(x,t+\xi)\,u(x,t-\xi)
\nonumber \\ && \hspace{1.5mm} {}
-u(x+\xi,t)\,u(x-\xi,t)\Big)\bigg|_{\xi=0} \,.
\end{eqnarray}

\subsection{Exact solutions}

\begin{figure}[t]
\begin{center}
\includegraphics[width=1.0\columnwidth]{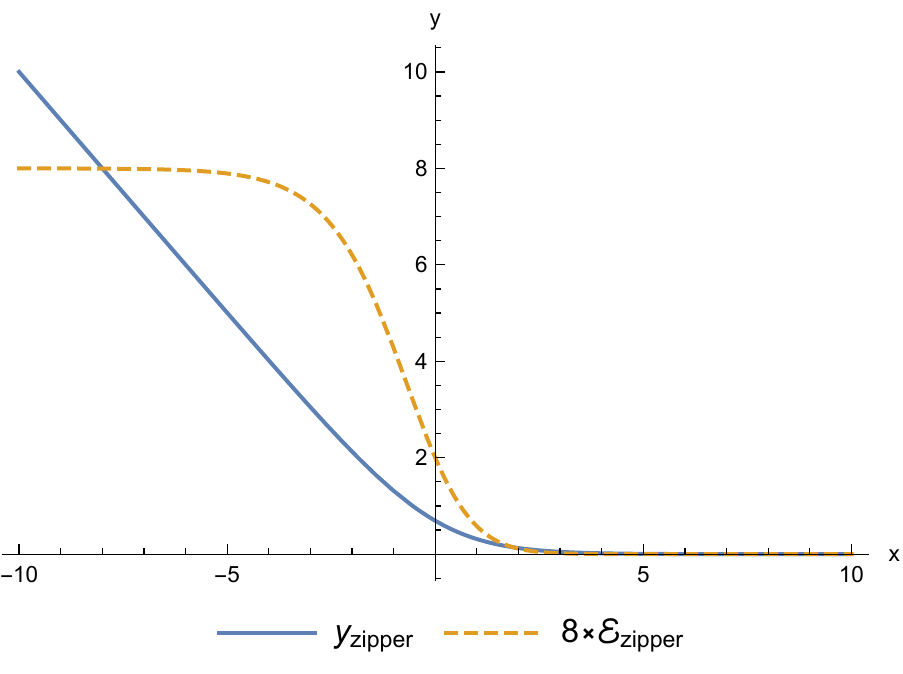}
\caption{\small A static zipper solution at the origin and its energy density (the graph is scaled for clarity).}
\label{fig:zipper}
\end{center}
\end{figure}

We can, in fact, find exact solutions of Eq.~\eqref{eq:u} using elementary guesswork inspired by the same techniques used in integrable systems. That is, putting a single exponential Ansatz $u = \e^{ k \cdot x}$, where $k \cdot x = k_t t - k_x x$, into \eqref{eq:u} gives us a condition $k^2 = -1$. Parametrising this as $k_t = v(1-v^2)^{-1/2}$ and $k_x = (1-v^2)^{-1/2}$ we get a moving solution:\footnote{The zipper solution and its stability was first investigated in \cite{Dauxois_2002}.}
\begin{equation}
y_{\mathrm{zipper}}(x) = \log\!\Big(1+\e^{-\tfrac{x-v t}{\sqrt{1-v^2}}}\Big) \,.
\end{equation}
This solution is very suggestive of a DNA molecule which is `unzipping' itself, as Fig.~\ref{fig:zipper} suggests. This solution is a soliton in the sense that it is an exact solution of a non-linear equation of motion which describes a  moving object. Furthermore, it is stable against perturbation as is shown in Appendix~\ref{app:A}. It has a semi-local nature, as its energy density approach $1$ on the far left, i.e., $\mathcal{E}_{\mathrm{zipper}} = (1+\e^{x})^{-2}$ which can be also seen in Fig.~\ref{fig:zipper}. The total energy of a zipper solution placed in the middle of a DNA strand of a length $L$ is
\begin{equation}
E_{\mathrm{zipper}} = \frac{1}{2}L-\tanh\bigg(\frac{L}{4}\bigg) \,.
\end{equation} 

\begin{figure}[t]
\begin{center}
\includegraphics[width=1.0\columnwidth]{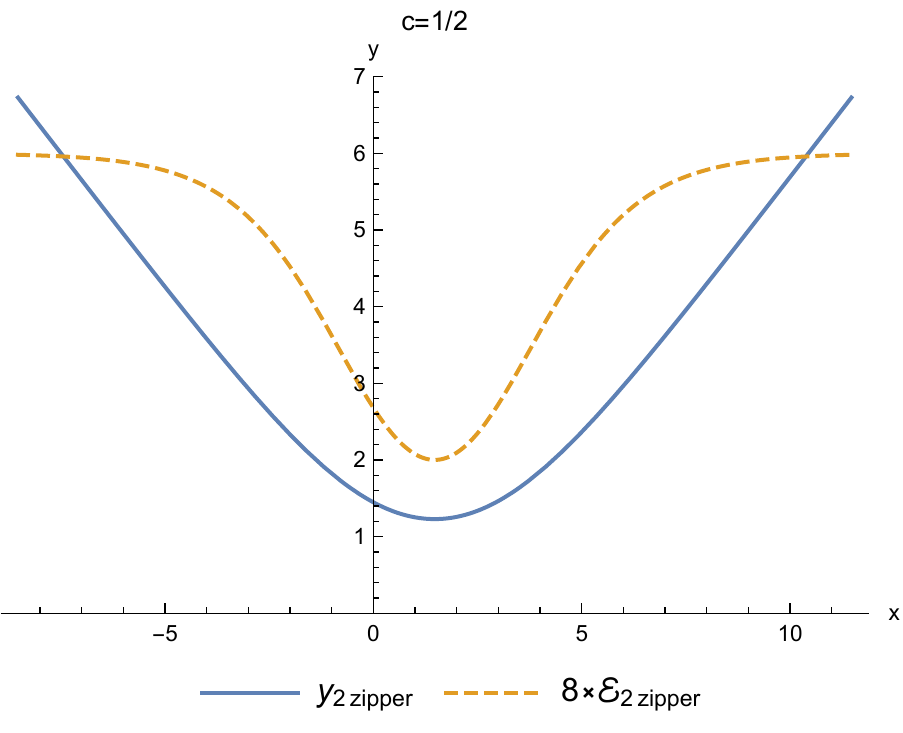}
\caption{\small A static two-zipper solution and its energy density for $c=1/2$ (the graph is scaled for clarity).}
\label{fig:twozipper}
\end{center}
\end{figure}

Experimenting with two-exponential Ansatz one can see that no new solutions arise. However, for three exponentials we obtain a new solution
\begin{equation}
y_{\mathrm{2\,zipper}}(x;c) = \log\frac{\e^{-x\sqrt{1-c}}+1+\frac{c}{4}\e^{x\sqrt{1-c}}}{1-c}\,,
\end{equation}
which reduces to a single zipper for $c=0$. Notice that this solution is defined only in the range $c \in [0,1)$. In fact, as Fig.~\ref{fig:twozipper} suggests, it describes a DNA which is `unzipped' from both sides centered at the point $x = \log(2/\sqrt{c}) / \sqrt{1-c}$. Let us stress, however, that we have not investigated the stability of this solution. Intuitively, it seems to represent an unstable equilibrium of equal and opposite unzipping of DNA  from both ends. At this point, we view it as a mere mathematical curiosity and whether it has any relevance for dynamics of DNA is a question left for a deeper investigation.    

Further experimentations with higher number of exponentials do not yield any new exact solutions. This suggests that Eq.~\ref{eq:u} is, in fact, \emph{not} integrable (as expected).


\subsection{Small amplitude perturbation expansion}

Let as now execute the algorithm of RG method. Putting in Eq.~\eqref{eq:u} a naive expansion
\begin{equation}
u = \varepsilon \big(u_0+\varepsilon u_1+\varepsilon^2 u_2 +\ldots\big) \,,
\end{equation}
we obtain a hierarchy of equations 
\begin{equation}
H_0 u_{n+1} = \sum_{k=0}^{n} \Big[(\partial_\mu u_{n-k})(\partial^\mu u_k)-u_{n-k}\partial^2 u_k\Big] \,,
\end{equation}
where $H_0 = \partial^2 + 1$. Starting with a monochromatic wave at the zero order, that is
\begin{equation}
u_0 = A_0 \e^{\I \theta} + \cc \,,
\hspace{3mm}
\theta \equiv q x-\omega t \,, \\
\end{equation}
where $\omega = \sqrt{q^2+1}$, the solution up to the second order reads
\begin{subequations}
\begin{eqnarray}
u_1 &=& 4 \abs{A_0}^2 \,, \\
u_2 &=& -2 A_0 \abs{A_0}^2 \e^{\I \theta} \Big(\xi \bar\theta^2 +\I (1-\xi)\theta\Big) + \cc
\end{eqnarray}
\end{subequations}
Here, $\xi$ is an arbitrary constant. We have also introduced an auxiliary variable
\begin{equation}
\bar\theta \equiv \omega x- q t \,,
\end{equation}
which appears in $u_2$ due to the identity\footnote{In fact, we will show in Appendix~\ref{app:B} that there are infinitely many possible secular terms, but that including these more complicated terms has no impact on the RG equation.}
\begin{equation}
\frac{1}{\partial^2 +1}\e^{\I \theta} = -\frac{1}{2}\e^{\I \theta} \Big[\xi\bar\theta^2+\I (1-\xi) \theta+c\Big]
\end{equation}
for an arbitrary constant $c$. 

Since we have two independent secular terms we introduce two renormalization scales $\theta_0$ and $\bar \theta_0$ as
\begin{equation}
\bar \theta^2 \longrightarrow \bar\theta^2 - \bar\theta_0^2 +\bar\theta_0^2 \,,
\hspace{5mm}
\theta \longrightarrow \theta-\theta_0 + \theta_0 \,.
\end{equation}
Then, we absorb the second $\bar\theta_0^2$ and $\theta_0$ by renormalizing the bare amplitude:
\begin{equation}
A_0 = \Big[1+2 \varepsilon^2\abs{A}^2 \big(\xi\bar \theta_0^2+\I (1-\xi) \theta_0\big)+ \mathcal{O}(\varepsilon^4) \Big]A \,,
\end{equation}
where $A \equiv A(\theta_0, \bar\theta_0)$ is the dressed amplitude. In this way, the solution becomes an explicit function of renormalization scales. However, these are unphysical parameters and we should demand that the full solution do not depend on them. Demanding that
\begin{equation}
\frac{\partial u}{\partial \theta_0} = \frac{\partial u}{\partial \bar\theta_0} = 0 \,,
\hspace{4mm}
\forall\, x,\,t \,,
\end{equation}
we obtain
\begin{subequations}
\begin{eqnarray}
\frac{\partial A}{\partial \theta_0} &=& -2\I \varepsilon^2 A \abs{A}^2 \big(1-\xi\big) + \mathcal{O}(\varepsilon^4) \,,
\\
\frac{\partial A}{\partial \bar\theta_0} &=& -2\varepsilon^2 A \abs{A}^2 \xi \bar \theta_0 + \mathcal{O}(\varepsilon^4) \,.
\end{eqnarray}
\end{subequations}
We cannot, however, take these equations at their face value, since 
RG equations must be fundamentally \emph{slow motion} equations, meaning that derivatives remains small for all values of $\theta_0$ and $\bar\theta_0$. However, this is clearly not the case since the right-hand side of the second equation is proportional to $\bar\theta_0$.\footnote{As far as we know, the condition that RG equations must be slow motion equation is not particularly stressed in the foundation papers \cite{Oono1, Oono2}. But it is a reasonable demand if we view it from the point of view of multiple-scales method (which is shown to be just a special case of RG method in \cite{Oono2}) as there it is self-evident. In this paper, we reinforce this condition by showing in App.~\ref{app:B} that the ambiguity in choosing the secular term is resolved precisely by ignoring those which does not lead to slow motion RG equations.}

To remedy this, instead of the second equation we should take its derivative, namely we get: 
\begin{eqnarray}
\frac{\partial A}{\partial \theta_0} &=& -2\I \varepsilon^2 A \abs{A}^2 \big(1-\xi\big) + \mathcal{O}(\varepsilon^4) \,,
\\
\frac{\partial^2 A}{\partial \bar\theta_0^2}  &=& - 4 \varepsilon^2 A \abs{A}^2 \xi+ \mathcal{O}(\varepsilon^4) \,.
\end{eqnarray}
We are still not finished, however, since the right-hand sides of both equation can become arbitrarily large if we take $\xi$ to be big. An obvious solution is to take an appropriate linear combination to arrive at the true RG equation which is a non-linear Schr\"odinger (NLS) equation:
\begin{equation}
\I \frac{\partial A}{\partial \theta_0} -\frac{1}{2}\frac{\partial^2 A}{\partial \bar\theta_0^2} = 2\varepsilon^2 A \abs{A}^2 \,.
\end{equation}

A single soliton solution of NLS equation is given as
\begin{equation}
\label{eq:nlssol}
A = \frac{\sigma}{\varepsilon\cosh\Big(\sigma\sqrt{2} \big(\bar\theta_0-v \theta_0\big)\Big)} \,\e^{-\I\Big(v \bar\theta_0+\big(\sigma^2 -\tfrac{v^2}{2}\big) \theta_0\Big)} \,,
\end{equation}
for arbitrary values of $\sigma$ and $v$. Notice that since in the full solution the amplitude appears only in combination $\varepsilon A$, the dependence on $\varepsilon$ completely disappears, as it should. However, since we are working with small amplitude expansion the above solution is reliable only for small values of $\sigma$.

Let us now plug the above into the solution and set $\theta_0 = \theta$ and $\bar\theta_0 = \bar\theta$ to eliminate secular terms. The renormalized solution thus obtained reads
\begin{eqnarray}
u_R^{I} &=& \frac{u_0}{\cosh\!\Big(\frac{u_0}{\sqrt{2}} \big(\bar\theta-v\theta\big)\Big)} \, \cos\!\Big(v\bar\theta+\big(u_0^2 -2v^2-4\big) \theta/4\Big)
\nonumber \\ && {}
+ \frac{u_0^2}{\cosh^2\!\Big(\frac{u_0}{\sqrt{2}} \big(\bar\theta-v \theta\big)\Big)} \,.
\label{eq:solsol2}
\end{eqnarray}
Here, we denoted $u_0 \equiv \sigma/2$ as a peak height of the soliton.

The soliton is moving with the envelope velocity
\begin{equation}
V_e \equiv \frac{q-v \omega}{\omega-v q} \,,
\end{equation}
which is nothing but relativistic addition of a group velocity $v_g = \frac{\d \omega}{\d q} = \frac{q}{\omega}$ with the amplitude velocity $-v$. 

\begin{figure}[t]
\begin{center}
\includegraphics[width=1.0\columnwidth]{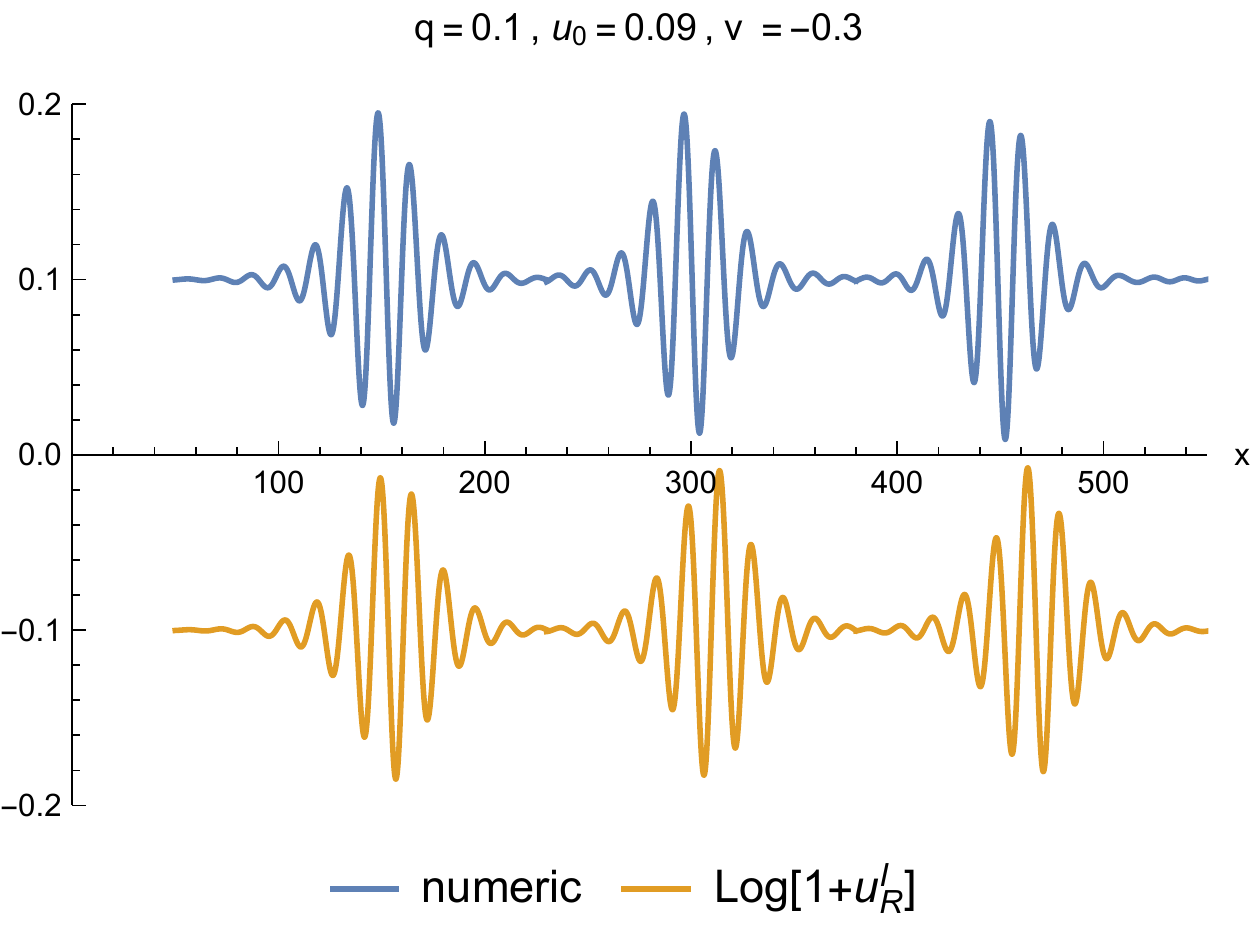}
\caption{\small Comparison between numerical solution of Eq.~\eqref{eq:y2} and renormalized solution \eqref{eq:solsol2}. The different time slices are taken (from the left) at $t=400$, $t=800$ and $t=1200$ time units. The graphs are offset from the horizontal axis for clarity. Notice that the bottom graphs deviates more and more from the upper ones as $t$ increases.}
\label{fig:02}
\end{center}
\end{figure}

However, comparing this solution with numerical calculations, we discover that it is not quite correct, as can be seen from Fig.~\ref{fig:02}. Namely, it seems that the numerical solution moves with different speed and oscillates at different frequency than our approximate solution.

\begin{figure}[t]
\begin{center}
\includegraphics[width=1.0\columnwidth]{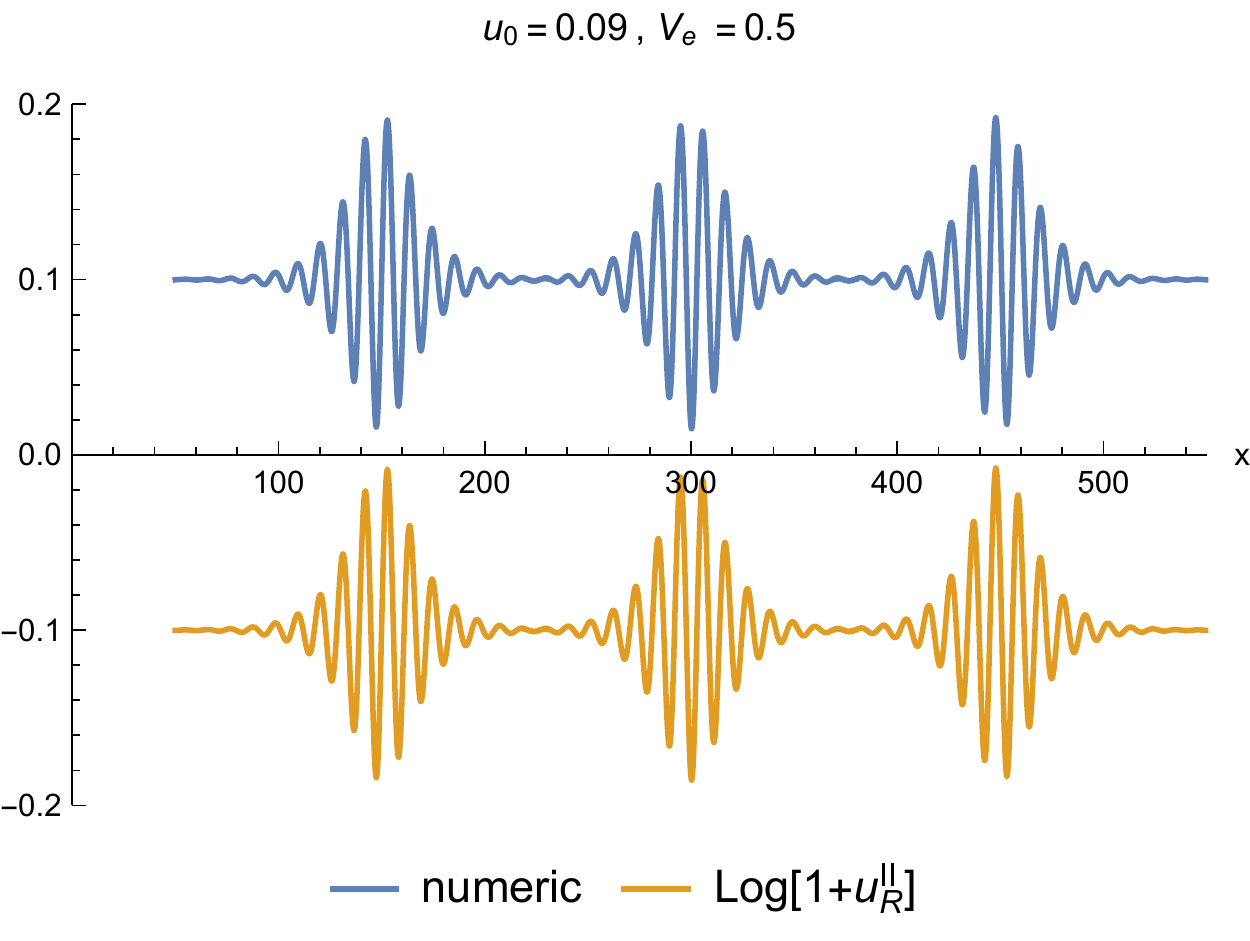}
\caption{\small Comparison between numerical solution of Eq.~\eqref{eq:y2} and renormalized solution \eqref{eq:solsol3}. The different time slices are taken (from the left) at $t=300$, $t=600$ and $t=900$ time units. The graphs are offset from the horizontal axis for clarity. Here we see persistent match between upper and bottom graphs for all range of $t$.}
\label{fig:03}
\end{center}
\end{figure}

Why is this happening? The culprit is the solution to NLS equation. We should be wary of its parameter $v$, which represent amplitude velocity and it is a manifestation of \emph{Galilean} invariance of NLS equation. But, our model respects Lorentzian symmetry, so we should not trust moving solutions of the NLS equation. Thus, to obtain a correct solution we should start from motionless NLS equation solution with $v=0$ and then boost it to the frame with the velocity
\begin{equation}
\frac{V_E-v_g}{1-V_Ev_g} \,,
\end{equation}
which is a relativistic sum of $V_E$, the final envelope velocity and the group velocity $v_g = q / \omega$. After the dust settles, we arrive at
\begin{eqnarray}
u_R^{II} &=& \frac{u_0}{\cosh\!\Big(\tfrac{u_0}{\sqrt{2}}\tfrac{x-V_E t}{\sqrt{1-V_E^2}}\Big)} \,\cos\!\bigg(\Big(1-\frac{u_0^2}{4}\Big)\frac{t-V_E x}{\sqrt{1-V_E^2}}\bigg)
\nonumber \\ && {}
+ \frac{u_0^2}{\cosh^2\!\Big(\tfrac{u_0}{\sqrt{2}}\tfrac{x-V_E t}{\sqrt{1-V_E^2}}\Big)} \,.
\label{eq:solsol3}
\end{eqnarray}
Notice that $q$ disappears! Comparing this solution with numerical solution in Fig.~\ref{fig:03} we indeed discover that now they match each other perfectly.

Aside from numerical simulation we can also support our claim 
by the following observation. If we plug these solutions into the full equation of motion \eqref{eq:u} and expand the results in terms of $u_0$, we learn that the generic solution \eqref{eq:solsol2} is $\mathcal{O}(u_0)$, while \eqref{eq:solsol3} is $\mathcal{O}(u_0^4)$! This disparity can be explained by observing that while the generic solution \eqref{eq:solsol2} has three free parameters, namely $q$, $V_E$ and $u_0$, the `relativistic' one only two, that is $V_E$ and $u_0$. In fact we can obtain \eqref{eq:solsol3} form \eqref{eq:solsol2} by fixing one of its parameters to eliminate  higher orders in $u_0$. Upon closely inspecting the expansion of Eq.~\eqref{eq:u} with the solution \eqref{eq:solsol2} inserted, one can immediately realize that first, second and third power of $u_0$ disappears if we set $v=0$ as we have argued. Then, the full transition from \eqref{eq:solsol2} to \eqref{eq:solsol3} is achieved by setting $q = V_E/\sqrt{1-V_E^2}$. In this way, we can be confident that the relativistic solution is indeed correct low-amplitude approximation.

Of course, the solution \eqref{eq:solsol2} is not only valid for $v = 0$ but also for velocities sufficiently small, namely 
if  $v \sim u_0^{3/4}$ then \eqref{eq:solsol2} is correct up to $\mathcal{O}(u_0^4)$.

Further notice, that for solution \eqref{eq:solsol3} the velocity of the envelope, that is $V_E$, is the same as the group velocity of the carrier wave. More precisely if we recast the argument of the cosine function as $\Theta x - \Omega t$, we see that $V_E = \Theta/\Omega$. In contrast, the so-called \emph{coherent mode} is achieved when the soliton moves with the phase velocity, i.e., $V_E = \Omega/ \Theta$, so that the internal oscillations match envelope velocity rendering the profile rigid. The idea of coherent mode has become popular in the literature \cite{Zdravkovic2, Slobodan} in order to reduce the parameter space of the approximate solution. Here, however, we see that hints coming from multiple directions all points towards a different scenario, where the soliton moves at (or very close to) the group velocity. These are i) comparison with numerical simulations, ii) analytic verification by plugging the solutions into the full equations of motion and, lastly, iii) the Lorentz invariance of the continuous PB model.

As we will see, the same conclusion is essentially reached in the discrete model, which we are going to tackle next.

\section{Solitons in the PB model}
\label{sec:V}

Let us now investigate solutions to PB model in the small amplitude limit, 
i.e., $a y_n \ll 1$. In other words, let us substitute $y_n = \varepsilon Y_n$, where $\varepsilon$ is a bookkeeping parameter, and let us expand the second equation of motion \eqref{eq:y} to the third order. We obtain
\begin{equation}
\label{eq:eomt}
H_0 Y_n = \varepsilon \alpha \omega_g^2 Y_n^2 -\varepsilon^2 \beta \omega_g^2 Y_n^3 \,, 
\end{equation}
with
\begin{equation}
\label{eq:mastereq}
\alpha \equiv \frac{3a}{\sqrt{2}} \,,
\hspace{5mm}
\beta \equiv \frac{7a^2}{3} \,,
\hspace{5mm}
\omega_g^2 \equiv \frac{4a^2 D}{m} \,,
\end{equation}
and where we collected all linear terms  into a single operator
\begin{equation}
\label{eq:h0}
H_0 \equiv \frac{\d^2}{\d t^2} -\frac{k}{m} \Big(\e^{\partial_n}+\e^{-\partial_n}-2\Big)+ \omega_g^2 \,.
\end{equation}

The RG method first calls for plugging a naive perturbation series
\begin{equation}
Y_n = Y_n^{(0)} + \varepsilon Y_n^{(1)} + \varepsilon^2 Y_n^{(2)} + \ldots
\end{equation}
into the truncated equation of motion \eqref{eq:eomt}. Solving it to the second order in $\varepsilon$, we get
\begin{subequations}
\begin{eqnarray}
Y_n^{(0)} &=& A_0 \e^{\I \theta_n} + \cc \,,
\\
Y_n^{(1)} &=& \frac{\alpha\,\omega_g^2}{\lambda_2} A_0^2 \e^{2\I \theta_n} + \alpha\abs{A_0}^2 + \cc \,,
\\
Y_n^{(2)} &=& \frac{\omega_g^2}{\lambda_2\lambda_3} \big(2\alpha^2\omega_g^2-\beta\lambda_2\big) A_0^3 \e^{3\I \theta_n}
\nonumber \\ && {}
-\frac{\omega_g^2}{2\omega \lambda_2} \big(2\alpha^2\omega_g^2+4\alpha^2\lambda_2-3\beta\lambda_2\big) A_0 \abs{A_0}^2
\nonumber \\ && \hspace{10mm} {}
\times 
\Big(\frac{\xi z^2}{V_g^{\prime}}-\I t(1-\xi)\Big) \e^{\I \theta_n} + \cc \,,
\end{eqnarray}
\end{subequations}
where
\begin{subequations}
\begin{eqnarray}
\theta_n &\equiv& nq l - \omega t \,,
\\
z &\equiv& n l- V_g t \,,
\\
\omega^2 &=& \omega_g^2 + \frac{4k}{m} \sin^2\!\Big(\frac{ql}{2}\Big)\,, \\
\lambda_p &=& \omega_g^2 + \frac{4k}{m}\sin^2\!\Big(\frac{pql}{2}\Big) - p^2 \omega^2 \,,
\\
V_g &\equiv& \frac{\d \omega}{\d q} = \frac{k l}{m \omega}\sin(q l) \,,
\\
V_g^{\prime} &\equiv& \frac{\d V_g}{\d q} = \frac{k l^2}{m\omega} \cos(q l)-\frac{V_g^2}{\omega}\,.
\end{eqnarray}
\end{subequations}
We have used the identity for the secular term in the form
\begin{equation}
\frac{1}{H_0} \e^{\I \theta_n} = - \frac{1}{2\omega} \Big(\frac{\xi z^2}{V_g^{\prime}}-\I t(1-\xi)\Big) \e^{\I \theta_n} \,,
\end{equation}
where $\xi$ is an arbitrary parameter.\footnote{Here, we assume that the ambiguity in choosing the secular terms is resolved in the similar manner as in the continuum PB model. See App.~\ref{app:B} for details.} 

\begin{figure}[t]
\begin{center}
\includegraphics[width=1.0\columnwidth]{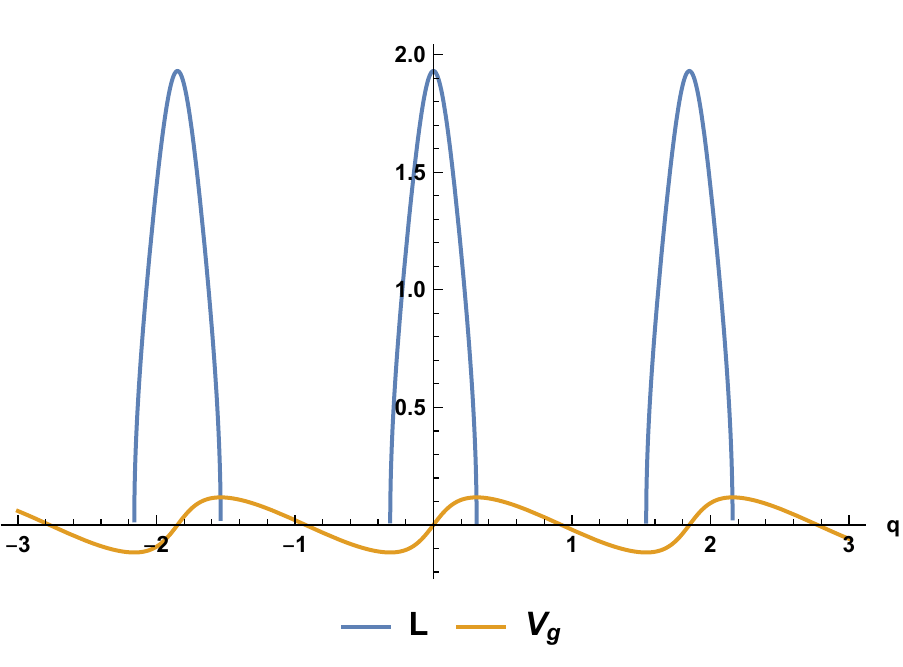}
\caption{\small The specific width of the soliton $L$ and the group velocity as a function of $q$. All other parameters takes their numerical values as defined in Sec.~\ref{sec:III}.}
\label{fig:04}
\end{center}
\end{figure}

Now we introduce renormalization scales $T, Z$ via
\begin{equation}
t \longrightarrow t-T+T \,,
\hspace{5mm}
z^2 \longrightarrow z^2 -Z^2 +Z^2 \,,
\end{equation}
and we absorb second $T$ and $Z^2$ into the definition of the dressed amplitude
\begin{multline}
A_0 = A \bigg[ 1 + \frac{\varepsilon^2\omega_g^2}{2\omega \lambda_2} \Big(2\alpha^2\omega_g^2+4\alpha^2\lambda_2-3\beta\lambda_2\Big) \abs{A}^2
\\
\times \Big(\frac{\xi Z^2}{V_g^{\prime}}-\I T(1-\xi)\Big) + \mathcal{O}(\varepsilon^4) \bigg] \,.
\end{multline}
The RG equations are given by demanding that the solution be independent of the renormalization scales, namely
\begin{equation}
\frac{\partial Y_n}{\partial T} = \frac{\partial Y_n}{\partial Z} = 0 \,,
\end{equation}
which yields
\begin{subequations}
\begin{eqnarray}
\I\frac{\partial A}{\partial T} &=& -(1-\xi)\varepsilon^2 Q A \abs{A}^2 + \mathcal{O}(\varepsilon^4) \,,
\\
\frac{\partial^2 A}{\partial Z^2} &=& -\xi \varepsilon^2 \frac{Q}{P} A \abs{A}^2Z + \mathcal{O}(\varepsilon^4) \,,
\end{eqnarray}
\end{subequations}
where
\begin{subequations}
\begin{eqnarray}
P &=& \frac{1}{2}V_g^{\prime} \,,
\\
Q &=& \frac{\omega_g^2}{2\omega \lambda_2} \Big(2\alpha^2\omega_g^2+4\alpha^2\lambda_2-3\beta\lambda_2\Big) \,.
\end{eqnarray}
\end{subequations}
In the same spirit as in the continuous limit we differentiate the second equation with respect to $Z$ and take a linear combination of the result with the first equation in such a way that we arrive at the $\xi$-independent, slow-motion RG equation, namely NLS equation:
\begin{equation}
\I \frac{\partial A}{\partial T} + P \frac{\partial^2 A}{\partial Z^2} + \varepsilon^2 Q A \abs{A}^2 = 0 \,.
\end{equation}
A soliton solution exists for $P/Q > 0$. In Fig.~\ref{fig:04} we plot the \emph{specific} width of the soliton $L \equiv \sqrt{2P/Q}$ and the group velocity $V_g$ to illustrate that this solution exists only for certain intervals of $q$ which are delimited by zeros of $P$. Their position can be found analytically as
\begin{equation}
q l = \pm \cos^{-1}\!\bigg[ 1 + \frac{2a^2 D}{k} \bigg(1- \sqrt{1+\frac{k}{a^2 D}}\bigg)\bigg] + 2\pi N \,,
\end{equation}
where $N \in \mathbb{Z}$.

Assuming that $q$ is always within these ranges a single soliton solution reads
\begin{equation}
A = \frac{\sigma\, \e^{\tfrac{\I v}{2P}\big(Z-v T/2\big)+\I \sigma^2 Q T/2}}{\varepsilon \cosh\!\Big(\sqrt{\tfrac{Q}{2P}}\sigma (Z-v T)\Big)} \,.
\end{equation}
Here, $\sigma$ manifest the scale invariance while $v$ represent Galilean boost invariance of NLS equation. Notice that since in the full solution the amplitude $A_n$ only occurs in the combination $\varepsilon A_n$, the bookkeeping parameter $\varepsilon$ cancels out everywhere.

\begin{figure}[t]
\begin{center}
\includegraphics[width=1.0\columnwidth]{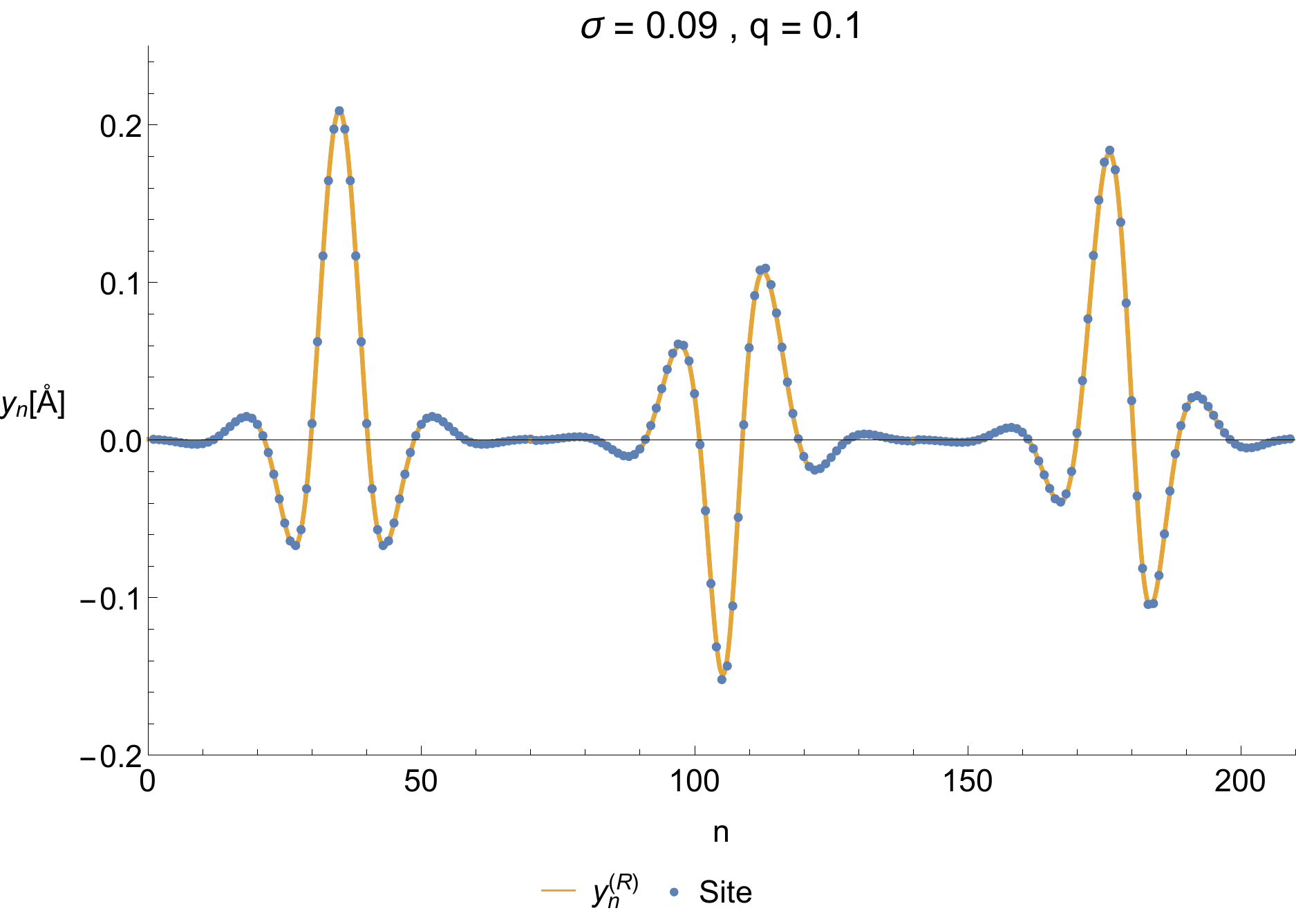}
\caption{\small The comparison between the renormalized solution of Eq.~\eqref{eq:renorsol} and a numerical solution in three different time stamps (ordered from left to right) with $\sigma = 0.09$ and $q=0.1$. The blue dots are positions of sites given by numerical calculations at $t= 0 \ \mathrm{t.u.}$ (first snapshot), $t = 80 \ \mathrm{t.u.}$ (second snapshot) and $t = 160 \ \mathrm{t.u.}$ (third snapshot), while the continuous yellow profile is given by the renormalized solution $y_n^{(R)}$. The vertical axis is shown in \aa ngstr\"oms.}
\label{fig:05}
\end{center}
\end{figure}

As was the case in the continuous limit, we should not expect that this solution remains valid for all values of $v$. In fact, plugging this solution into the perturbative expansion for $y_n$ and putting the whole into equation of motion \eqref{eq:eomt} one can check that the solution will be correct up to third order in $\sigma$, if one takes $v=0$. As before, for sufficiently small $v$ the solution can also be correct up to third order. Here, however, we must demand $v \sim \sigma^2$, unlike in the continuous limit. For $v=0$ the renormalized solution reads
\begin{eqnarray}
y_n^{(R)} &=& 
\frac{2\sigma \cos\!\big(n l \Theta -\Omega t\big)} {\cosh\!\Big(\tfrac{\sigma}{L}\big(n l -V_g t\big)\Big)}
\nonumber \\ && {}
+ \frac{2\alpha \sigma^2}{\lambda_2}\, \frac{\omega_g^2 \cos\!\big(2n l \Theta -2\Omega t\big) +  \lambda_2}{\cosh^2\!\Big(\tfrac{\sigma}{L}\big(n l -V_g t\big)\Big)}
\nonumber \\ && {}
+\frac{2\omega_g^2 \sigma^3(2\alpha^2 \omega_g^2-\beta \lambda_2)}{\lambda_2\lambda_3} \, \frac{\cos\!\big(3n l \Theta -3\Omega t\big)}{\cosh^3\!\Big(\tfrac{\sigma}{L}\big(n l -V_g t\big)\Big)} \,,
\nonumber \\ && \label{eq:renorsol}
\end{eqnarray}
where
\begin{equation}
\Theta = q \,,
\hspace{6mm}
\Omega = \omega -\frac{\sigma^2 Q}{2} \,,
\hspace{6mm}
L = \sqrt{\frac{2P}{Q}} \,.
\end{equation}
We compare this solution to the numerical solution of Eq.~\eqref{eq:y} in Fig.~\ref{fig:05} for $\sigma =0.09$ and $q = 0.1$ which corresponds to $V_g \approx 0.07$. We observe a near perfect-match between the two for a long interval of time. Our numerical experiments confirms this result for various values of  $\sigma$ and $V_g$ reinforcing our belief that $y_n^{(R)}$ is a correct, uniform, small-scale approximation to the true solution of the full equation of motion Eq.~\eqref{eq:y}. 

The specific width of the soliton $L$ has a maximum in the middle of intervals of allowed values of $q$, namely at $q = 2N\pi/l$. Notice that at these values the soliton is at rest, i.e., $V_g = 0$, while the velocity is largest in the absolute sense at the edges of the allowed intervals for $q$. At these edges, however, the width of the soliton is zero and we obtain a degenerate case $y_n = 0$.

The true maximum width of the soliton $W_{\mathrm{max}}$, which we defined as the interval in which half of the area of the envelope $2\sigma/\cosh(\sigma x/L)$ is contained, reads
\begin{equation}
W_{\mathrm{max}}
\ \approx \ 
1.76 \, \frac{L}{\sigma} 
\ = \ 
1.76 \, \frac{l\sqrt{k}}{4a^2 \sigma \sqrt{D}} 
\ \approx \ 
\frac{3.4}{\sigma} \, [\angstrom]\,.
\end{equation}
In \cite{Slobodan} it is argued that the typical size of DNA segment participating in the transcription process is between 8 to 17 nucleotides. Thus, if we want the width of the soliton to be roughly the same, we obtain an interval for the value of $\sigma$ as
\begin{equation}
8 \ \leq \ \frac{W_{\mathrm{max}}}{l} \ \leq \  17 \,,
\hspace{5mm} \Rightarrow \hspace{5mm}
1/17 \ < \  \sigma \ < \ 1/8 \,,
\hspace{4mm}
\end{equation}
which is comfortably small enough that we can trust our approximate solution to be applicable for such case.
 
\section{Discussion}
\label{sec:VI}

We have derived an approximate solitonic solution of the Peyrard--Bishop model using the Renormalization Group perturbation expansion method developed by Chen, Goldenfeld and Oono \cite{Oono1}. Compared to traditional multiple-scale analysis, which is used almost exclusively throughout the literature,
this method has several advantages. First, it is a straightforward algorithm. There is no need to make guesses about the nature of the slow scales \emph{a priori} nor it is necessary to justify them \emph{a posteriori}. The derivation of the approximate solution follows naturally from the logic of the method and all new parameters, which pop up during its execution, arise with a clear physical meaning.

Second, the RG method is systematic, therefore higher-order corrections can be carried out directly. In contrast, higher-order corrections in the multiple-scale method require quite an art to calculate, as new scales may have to be included. In this way, RG method is well-suited for the exploration of higher-order effects of solitons in the dynamics of DNA.

Thirdly, the RG method does not require continuous limit as an additional approximation, as it is the case for the multiple-scale analysis when applied to PB model. While quite natural, this simplification strains the applicability of the resulting solution. For example, if the soliton covers only $10-20$  nucleotides (which is the size of the denaturation bubble and, correspondingly, a probable size of the soliton), one may not be quite comfortable with the idea of a continuum. In comparison, no such difficulty arises in RG method, where the solution's validity depends only on the size of its amplitude and not on its width. This is so because the effective RG equations which govern the \emph{artificial} renormalization scales (as opposed to physical ones in multiple-scale method) are automatically continuous even if the method is used in discrete models.

In this work, we have also claimed that the optimal velocity of the soliton's envelope -- as far as the approximate dynamics investigated here is to be believed -- should be the \emph{group velocity} of the carrier wave. This is in conflict with the proposal that the soliton should be in the so-called \emph{coherent mode} \cite{Zdravkovic2}. In the coherent mode, the envelope would travel at \emph{phase velocity} of the carrier wave, hence the soliton's profile would not change over time. However, comparing our results with numerical simulations and via directly plugging of the approximate solution into the equations of motion, we have found that it is when the envelope travels with group velocity (or velocity very close to it) that the solution is reliable. Furthermore, this observation can be explained in the continuous PB model as a consequence of Lorentz invariance of the model, which forbids the Galilean boost symmetry of the non-linear Schr\"odinger equation (the RG equation for the envelope). Let us, however, state that we do not claim that our observation disproofs the idea of the coherent mode, which may be valid for other reasons. We only claim that the coherent mode does not arise as the most reliable approximate solution of the PB model.

Let us also point out that we do not claim that the solution \eqref{eq:renorsol} is equally physically relevant for the entire allowed range of parameters $\sigma$ and $V_g$. For instance, the non-moving solution $V_g =0$ imply that solitons can remain fixed at a particular place on DNA. This seems to be aligned with the idea that solitons contributes to the formation of  local opening of DNA. However, to decide whether that is true or, in general, what is the range of parameters that can be realistically applied for DNA physics is beyond the ambition of this paper.

Lastly, let us stress that the RG method can be readily used for more complicated models, such as Peyrard--Bishop--Dauxois model or others. Given that this method is both technically and  conceptually easier than the multiple-scale method, it would be interesting to explore solitonic solutions of other PB-like models and higher-order corrections therein to fully explore the idea of solitons in mechanical models of DNA.

\acknowledgments

This work was supported by the program of Czech Ministry of Education, Youth and Sports INTEREXCELLENCE Grant number LTT17018 (F.~B., P.~B.). P.~B. also thanks TJ~Balvan Praha for support.

\appendix 

\section{Stability of the zipper}
\label{app:A}

\begin{figure}[t]
\begin{center}
\centering
\includegraphics[width= 1.0\columnwidth]{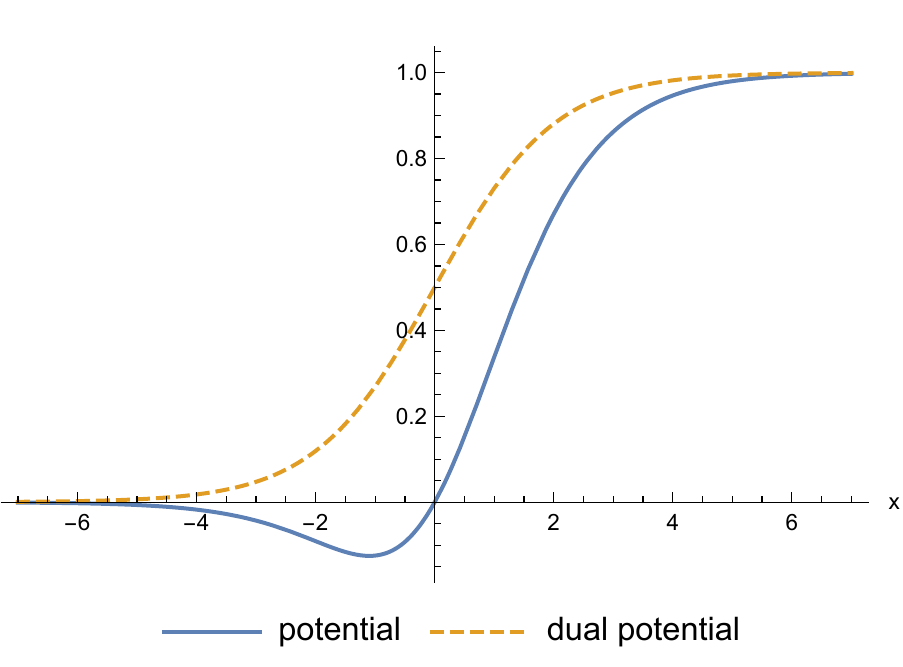}
\caption{\small Schr\"odinger potential for the zipper and its dual.}
\label{fig:potential}
\end{center}
\end{figure}

Let us study the spectrum of small fluctuations around the (static) zipper. For that end we set
\begin{equation}
y = y_{\mathrm{zipper}} + \e^{\I \omega t} \, b(x) \,,
\end{equation}
and linearize full equation of motion assuming $\abs{b(x)} \ll 1$. In this way, we obtain a Schr\"odinger-like eigenproblem
\begin{equation}
-b^{\prime\prime}(x) + \frac{\e^{x}\big(\e^{x}-1\big)}{\big(\e^{x}+1\big)^2}b(x) = \omega^2 \, b(x) \,.
\end{equation}
The potential is shown in Fig.~\ref{fig:potential}.

The translational zero mode $b_0(x) = \e^{-x} / (1+\e^{-x})$ is not normalizable (as to be expected for a divergent solution). The second independent solution reads
\begin{equation}
\tilde b_0(x)
\propto
b_0(x) \int^x \! \frac{\d x^{\prime}}{b_0^2(x^\prime)}
=
2\partial_c y_{\mathrm{2\,zipper}}(x,0)-2 b_0(x)
\end{equation}
and is not normalizable either.

However, from Fig.~\ref{fig:potential} we see that there is a small potential well which suggests a possibility of a normalizable bound state with a \emph{negative energy}, which would point to instability. Utilizing methods of supersymmetric quantum mechanics we can find the Hamiltonian and its dual in the form
\begin{subequations}
\begin{eqnarray}
H &=& \big(-\partial_x +W\big)\big(\partial_x +W\big) =  -\partial_x^2 +\frac{1-\e^{-x}}{\big(1+\e^{-x}\big)^2} \,,
\\
\tilde H &=& \big(\partial_x +W\big)\big(-\partial_x +W\big) =  -\partial_x^2 +\frac{1}{1+\e^{-x}} \,,
\hspace{15mm}
\end{eqnarray}
\end{subequations}
where
\begin{equation}
W = \frac{1}{1+\e^{-x}} \,.
\end{equation}
Fortunately, the dual is equivalent to a special case of the Rosen--Morse II potential, which is exactly solvable. The eigenfunctions and the eigenvalues ($\tilde H\tilde b_n = \tilde E_n \tilde b_n$) respectively read
\begin{eqnarray}
\tilde b_n &=& \e^{\tfrac{x}{2(1+n)}} \cosh^{1+n}\!\Big(\frac{x}{2}\Big) \, P_{n}^{(s_1,s_2)}\! \bigg(\tanh\!\Big(\frac{x}{2}\Big)\bigg) \,,
\hspace{8mm}
\\
\tilde E_n &=& -\frac{1}{4} s_2^2 \ = \  - \frac{1}{4}\big[(1+n)^2+(1+n)^{-2}-2\big] \,,
\end{eqnarray}
where $s_1 = -(1+n)-(1+n)^{-1}$, $s_2 = (1+n)^{-1}-(1+n)$ and where $P_n^{(s_1,s_2)}(x)$ are Jacobi polynomials. Obviously, none of the above eigenfunctions are normalizable states and they are solutions of the eigenproblem only in a formal sense. Also notice that the tower of eigenstates is in the negative energy direction as there are no discrete states for positive energy.

The energy of the normalized bound state must be above the potential minimum which is $-1/8$. However, since the first state $E_1 = -9/16$ is already under this value, we can safely conclude that 
no tachyonic state exists and the zipper solution is, therefore, stable.

\section{Ambiguity of a secular term in the continuous model}
\label{app:B}

The secular term in the continuous PB model arise as a solution to
\begin{equation}
H_0 \Big(f(\theta, \bar\theta)\e^{\I \theta}\Big) = \e^{\I \theta} \,,
\end{equation}
where $H_0 \equiv \partial_\theta^2 - \partial_{\bar\theta}^2 + 1$ is the zero-order operator and where we used convenient coordinates $\theta \equiv q x-\omega t$, $\bar\theta \equiv \omega x-q t$ with $\omega^2 = q^2+1$. As this represents a partial differential equation, we expect that $f(\theta,\bar\theta)$ is determined only up to two arbitrary functions, representing `initial' conditions, i.e., $f(\theta,0) \equiv g_0(\theta)$ and $\partial_{\bar\theta}f(\theta,0) \equiv g_1(\theta)$.

Indeed, a general solution reads
\begin{eqnarray}
f(\theta,\bar\theta) &=& 
- \frac{1}{2} \bar\theta^2 + \cosh\!\Big(\bar\theta\sqrt{\partial_\theta^2+2\I \partial_\theta}\Big) g_0(\theta)
\nonumber \\ && \hspace{8.5mm} {}
+ \frac{\sinh\!\Big(\bar\theta\sqrt{\partial_\theta^2+2\I \partial_\theta}\Big)}{\sqrt{\partial_\theta^2+2\I \partial_\theta}} g_1(\theta) \,.
\end{eqnarray}
Taking $g_0 = -\I (1-\xi)\theta/2$ and $g_1 =0$, we obtain the standard form $f = -\bar\theta^2 \xi/2-\I (1-\xi)\theta/2$. Alternatively, let us take $g_0 = 0$ and $g_1 = \xi \theta$, which leads to $f = -\bar\theta^2/2 +\xi \theta \bar\theta + \I \xi \bar\theta^3/3$. Such a choice would lead to RG equations in the form
\begin{subequations}
\begin{eqnarray}
\frac{\partial A}{\partial \theta_0} &=& 4 \varepsilon^2 A \abs{A}^2 \xi \bar\theta_0 + \mathcal{O}(\varepsilon^4) \,,
\\
\frac{\partial A}{\partial \bar\theta_0} &=& 4\varepsilon^2 A \abs{A}^2 \big(-\bar\theta_0 +\xi \theta_0 + \xi \I \bar\theta_0^2\big) + \mathcal{O}(\varepsilon^4) \,.
\hspace{10mm}
\end{eqnarray}
\end{subequations}
The slow-motion $\xi$-independent equation is obtained by taking a combination
\begin{equation}
2\I \frac{\partial A}{\partial \theta_0} - \frac{\partial^2 A}{\partial \bar\theta_0^2} = 4 \varepsilon^2 A \abs{A}^2 + \mathcal{O}(\varepsilon^4) \,.
\end{equation}
Indeed, this is exactly the same NLS equation obtained in the main text. 

Let us now consider $g_0 = \xi \theta^2$ and $g_1 = 0$ giving us $f = (\xi -1/2)\bar\theta^2 + \xi \theta^2 + 2\I \xi \theta \bar \theta^2 - \frac{1}{3} \xi \bar\theta^4$. This choice leads to
\begin{subequations}
\begin{eqnarray}
\frac{\partial A}{\partial \theta_0} &=& 4 \varepsilon^2 A \abs{A}^2 \big(2\xi \theta_0+2\I \xi \bar\theta_0^2\big) + \mathcal{O}(\varepsilon^4) \,,
\\
\frac{\partial A}{\partial \bar\theta_0} &=& 4\varepsilon^2 A \abs{A}^2 \bigg((2\xi-1)\bar\theta_0+4\I \xi \theta_0\bar\theta_0 - \frac{4}{3}\xi \bar\theta_0^3\bigg) + \mathcal{O}(\varepsilon^4) \,.
\nonumber \\ &&
\end{eqnarray}
\end{subequations}
Here, we are faced with a puzzle, since there exist two valid  combinations, namely
\begin{eqnarray}
2\I \frac{\partial A}{\partial \theta_0} + \frac{\partial^2 A}{\partial \theta_0^2} - \frac{\partial^2 A}{\partial \bar\theta_0^2} &=& 4 \varepsilon^2 A \abs{A}^2 + \mathcal{O}(\varepsilon^4) \,.
\\
2\I \frac{\partial^2 A}{\partial \theta_0\partial \bar\theta_0} &=& \frac{\partial^3 A}{\partial \bar\theta_0^3} + \mathcal{O}(\varepsilon^4) \,.
\end{eqnarray}
None of these, however, is a slow-motion equation. The second one is a differential consequence of NLS equation (up to $\mathcal{O}(\varepsilon^4)$), but it is not slow-motion. It is, however, suspicious as it does not contain explicit dependence of $\varepsilon$. The first of these appears to be a valid slow-motion equation at a first sight, however, it is a \emph{hyperbolic} equation and we can argue that due to the term $\partial_{\theta_0}^2A -\partial_{\bar \theta_0}^2A$ the second derivatives are not under control and can get very large. Following the policy of disregarding RG equations which are not slow-motion we must, therefore, ignore the corresponding secular terms.

For general $f$, the RG equations read
\begin{subequations}
\begin{eqnarray}
\frac{\partial A}{\partial \theta_0} &=& 4 \varepsilon^2 A \abs{A}^2 \partial_{\theta_0}\,f(\theta_0, \bar\theta_0) + \mathcal{O}(\varepsilon^4) \,,
\\
\frac{\partial A}{\partial \bar\theta_0} &=& 4\varepsilon^2 A \abs{A}^2 \partial_{\bar \theta_0}\,f(\theta_0, \bar\theta_0) + \mathcal{O}(\varepsilon^4) \,.
\end{eqnarray}
\end{subequations}
Given that $\partial_\theta^2f -\partial_{\bar\theta}^2f +2\I \partial_\theta f = 1$, we can always combine the above into
\begin{equation}
2\I \frac{\partial A}{\partial \theta_0}+\frac{\partial^2 A}{\partial \theta_0^2}-\frac{\partial^2 A}{\partial \bar\theta_0^2} = 4 \varepsilon^2 A \abs{A}^2 + \mathcal{O}(\varepsilon^4) \,,
\end{equation}
which is not slow-motion. Only if $\partial_\theta^2f = 0$ we obtain the correct slow-motion RG equation, which is the NLS equation: 
\begin{equation}
2\I \frac{\partial A}{\partial \theta_0}-\frac{\partial^2 A}{\partial \bar\theta_0^2} = 4 \varepsilon^2 A \abs{A}^2 + \mathcal{O}(\varepsilon^4) \,.
\end{equation}


\providecommand{\href}[2]{#2}\begingroup\raggedright\endgroup

\end{document}